\documentclass[aps,pre,a4paper,10pt,twocolumn,superscriptaddress]{revtex4-1} 
\usepackage{amsmath} 
\usepackage{amssymb} 
\usepackage{bbm} 
\usepackage{graphicx} 
\usepackage{epstopdf} 
\usepackage{xcolor}
\usepackage{hyperref} 
\usepackage{comment}

\begin{document} 
%%%%%%%%%%%%%%%%%%%%%%%%%%%%%%%%%%%%%%%%%%%%%%%%%%%%%%%%%%%%%%%%%%%%%%%%%%%%%%%%%%%%%%%%%%%%%%%%%%%%%%%%%%% 
%%%%%%%%%%%%%%%%%%%%%%%%%%%%%%%%%%%%%%%%%%%%%%%%%%%%%%%%%%%%%%%%%%%%%%%%%%%%%%%%%%%%%%%%%%%%%%%%%%%%%%%%%%% 
%%%%%%%%%%%%%%%%%%%%%%%%%%%%%%%%%%%%%%%%%%%%%%%%%%%%%%%%%%%%%%%%%%%%%%%%%%%%%%%%%%%%%%%%%%%%%%%%%%%%%%%%%%% 
%%%%%%%%%%%%%%%%%%%%%%%%%%%%%%%%%%%%%%%%%%%%%%%%%%%%%%%%%%%%%%%%%%%%%%%%%%%%%%%%%%%%%%%%%%%%%%%%%%%%%%%%%%% 
\title{Brownian motion surviving in the unstable cubic potential\\ and the role of Maxwell's demon} 
%%%%%%%%%%%%%%%%%%%%%%%%%%%%%%%%%%%%%%%%%%%%%%%%%%%%%%%%%%%%%%%%%%%%%%%%%%%%%%%%%%%%%%%%%%%%%%%%%%%%%%%%%%% 
%%%%%%%%%%%%%%%%%%%%%%%%%%%%%%%%%%%%%%%%%%%%%%%%%%%%%%%%%%%%%%%%%%%%%%%%%%%%%%%%%%%%%%%%%%%%%%%%%%%%%%%%%%% 
\author{Luca Ornigotti} 
\email{luca.ornigotti@gmail.com} 
\affiliation{Palack{\' y} University, Department of Optics, 17.~listopadu~1192/12, 771~46~Olomouc, Czech Republic} 

\author{Artem Ryabov} 
\email{rjabov.a@gmail.com} 
\affiliation{Charles University, Faculty of Mathematics and Physics, Department of Macromolecular Physics, V Hole{\v s}ovi{\v c}k{\' a}ch~2, 180~00~Praha~8, Czech Republic} 

\author{Viktor Holubec} 
\email{Viktor.Holubec@mff.cuni.cz} 
\affiliation{Charles University, Faculty of Mathematics and Physics, Department of Macromolecular Physics, V Hole{\v s}ovi{\v c}k{\' a}ch~2, 180~00~Praha~8, Czech Republic} 
\affiliation{
Universit{\"a}t Leipzig,
Institut f{\"u}r Theoretische Physik,
Postfach 100 920, D-04009 Leipzig, Germany
}

\author{Radim Filip} 
\email{filip@optics.upol.cz} 
\affiliation{Palack{\' y} University, Department of Optics, 17.~listopadu~1192/12, 771~46~Olomouc, Czech Republic}

%%%%%%%%%%%%%%%%%%%%%%%%%%%%%%%%%%%%%%%%%%%%%%%%%%%%%%%%%%%%%%%%%%%%%%%%%%%%%%%%%%%%%%%%%%%%%%%%%%%%%%%%%%%
%%%%%%%%%%%%%%%%%%%%%%%%%%%%%%%%%%%%%%%%%%%%%%%%%%%%%%%%%%%%%%%%%%%%%%%%%%%%%%%%%%%%%%%%%%%%%%%%%%%%%%%%%%%
\date{\today} 
%%%%%%%%%%%%%%%%%%%%%%%%%%%%%%%%%%%%%%%%%%%%%%%%%%%%%%%%%%%%%%%%%%%%%%%%%%%%%%%%%%%%%%%%%%%%%%%%%%%%%%%%%%%
%%%%%%%%%%%%%%%%%%%%%%%%%%%%%%%%%%%%%%%%%%%%%%%%%%%%%%%%%%%%%%%%%%%%%%%%%%%%%%%%%%%%%%%%%%%%%%%%%%%%%%%%%%%
\begin{abstract} 
Trajectories of an overdamped particle in a highly unstable potential diverge so rapidly, that the variance of position grows much faster than its mean. Description of the dynamics by moments is therefore not informative. Instead, we propose and analyze local directly measurable characteristics, which overcome this limitation. We discuss the most probable particle position (position of the maximum of the probability density) and the local uncertainty in an unstable cubic potential, $V(x)\sim x^{3}$, both in the transient regime and in the long-time limit. The maximum shifts against the acting force as a function of time and temperature. Simultaneously, the local uncertainty does not increase faster than the observable shift. In the long-time limit, the probability density naturally attains a quasi-stationary form. We interpret this process as a stabilization via the measurement-feedback mechanism, the Maxwell demon, which works as an entropy pump. Rules for measurement and feedback naturally arise from basic properties of the unstable dynamics. All reported effects are inherent in any unstable system. Their detailed understanding will stimulate the development of stochastic engines and amplifiers and later, their quantum counterparts.
\end{abstract} 
%%%%%%%%%%%%%%%%%%%%%%%%%%%%%%%%%%%%%%%%%%%%%%%%%%%%%%%%%%%%%%%%%%%%%%%%%%%%%%%%%%%%%%%%%%%%%%%%%%%%%%%%%%%
%%%%%%%%%%%%%%%%%%%%%%%%%%%%%%%%%%%%%%%%%%%%%%%%%%%%%%%%%%%%%%%%%%%%%%%%%%%%%%%%%%%%%%%%%%%%%%%%%%%%%%%%%%%
% insert suggested PACS numbers in braces on next line
% \pacs{}
% insert suggested keywords - APS authors don't need to do this
%\keywords{}
%\maketitle must follow title, authors, abstract, \pacs, and \keywords
\maketitle
%%%%%%%%%%%%%%%%%%%%%%%%%%%%%%%%%%%%%%%%%%%%%%%%%%%%%%%%%%%%%%%%%%%%%%%%%%%%%%%%%%%%%%%%%%%%%%%%%%%%%%%%%%%
%%%%%%%%%%%%%%%%%%%%%%%%%%%%%%%%%%%%%%%%%%%%%%%%%%%%%%%%%%%%%%%%%%%%%%%%%%%%%%%%%%%%%%%%%%%%%%%%%%%%%%%%%%%
\section{Introduction}
%%%%%%%%% 

Various engines and amplifiers exploit a natural instability in their parts to perform useful work or required manipulations. Instability is therefore a resource, although, it is simultaneously dangerous for the system. It can, in fact prevent the machine from working or, in a drastic case, it can completely damage it. Unstable systems, when left to evolve freely, have a strong tendency to diverge during quite a short period of time. Their variables can reach unwanted extremely large values for finite time intervals. Speaking statistically, not only all their statistical moments diverge, but, even more destructively, standard deviations can diverge faster than mean values. At this moment, the moments cannot inform about the unstable stochastic dynamics, and a different approach is required. An illustrative example of such instability is the unbounded cubic potential, 
\begin{equation}
\label{potential}
V(x)= \frac{1}{3} k x^3 , 
\end{equation} 
which exhibits all these aspects even in an overdamped regime.

Recently, the investigation of unstable systems got a large stimulus from the experimental development. Beyond the overdamped regime, the cubic nonlinear potential is experimentally accessible in the developing field of optomechanics with both nanoparticles \cite{Gieseler2013, GieselerQuidantPRL2014, GieselerNovotnyNJP2015, NovotnyQuidant2017} and solid-state objects \cite{HarrisPRA2012, Schaefermeier2016}. In quantum optomechanics, the cubic nonlinearity is principally required to construct highly nonlinear Hamiltonians and potentially, implement analog quantum simulations with mechanical objects \cite{LloydPRL99, GottesmanPRA2001, MiyataPRA2016}. Investigation of unstable systems is also important for a development of quantum mechanical engines beyond simple double-well models \cite{GavrilovPRL2016, 2017arXiv170511180G}, which is necessary for further development of quantum thermodynamics. All these investigations also require both comparison with and understanding of the overdamped case.

In the present work, we thus discuss dynamics of an overdamped Brownian particle diffusing in the unstable cubic potential~\eqref{potential}. Even though we focus on the particular case of a cubic potential, our approach can be easily generalized to other unstable potentials with an inflection point. We assume that position of the particle $x(t)$ evolves in time according to the Langevin equation
\begin{equation}
\label{Langevin}
 \frac{dx}{dt} = - \frac{k}{\gamma} x^2(t) + \sqrt{2 D} \xi (t),
\end{equation}
where $\xi(t)$ is the standard Gaussian white noise [$\langle \xi(t) \rangle=0$, $\langle \xi(t)\xi(t') \rangle=\delta(t-t')$], $\gamma$ stands for the friction and the diffusion coefficient satisfies the fluctuation-dissipation theorem, $D=k_{\rm B} T/\gamma$. Though all derivations are carried out for arbitrary $\gamma$, in illustrations we will always assume $\gamma=1$.

Brownian dynamics described by Eq.~\eqref{Langevin} occurs as a basic element of several nonlinear stochastic models in chemistry, physics and biology. Examples include firing of noisy neurons \cite{Lindner2003, Brunel2003}, optical bistability in lasers \cite{SanchoPRA1989, SanchoPRA1991, ArecchiPRA1971}, or, more generally, passage through the saddle-node bifurcation \cite{Sigeti1989, HirschPRA1982, ReimannBroeck94}, where the simplicity of Eq.~\eqref{Langevin} allows to describe a phenomenon of intermittency. Another broad class of systems where Eq.~\eqref{Langevin}  occurs naturally, are one-dimensional Brownian ratchets modeled as diffusion in tilted-periodic potentials. Transport properties of the latter at a critical tilt were derived in Refs.~\cite{ReimannPRL01, ReimannPRE02, Dean1}. Other examples of transitions from metastable states in condensed matter models can be found in Refs.~\cite{TretiakovPRB2003, SpagnoloEntropy2017}.

Analytical techniques developed to describe such problems (decays of unstable states) can be roughly subdivided into three groups. The first deals with first-passage times \cite{bookRedner, HanggiRevModPhys1990, Arecchi1982, YoungPRA1985, SanchoPRA1989, HirschPRA1982, Sigeti1989, SanchoPRA1991, Caceres1995, MantegnaPRL1996, AgudovPRE1998, Lindner2003, Brunel2003, FiasconaroPRE2005, CaceresJSP2008, ARPZRF2016, SciRep2017}. 
The second, a rather related one, focuses on a so called nonlinear relaxation time (or a mean time spent by a particle in a given region) \cite{AgudovMalakhovPRE1999, DubkovPRE2004}. This approach differs from the first-passage approach by accounting for multiple passages (returns) of the particle and not only for the first one. 
Third, significant effort was made to analyze time-evolution of the probability density function (PDF) in a symmetric inverted parabolic potential (bounded by a quartic potential for large $x$) \cite{Suzuki1976I, Suzuki1976II, Suzuki1977III, HaakePRL1978, PasqualeTombesi1979, DEKKER1980, PasqualePRA1982, WeissPRA1982, DEKKER1982}. 
Whereas the first two approaches provide only indirect information about the particle position, the third aims directly at the position PDF. It exploits symmetry of the problem and/or properties of the inverted parabolic potential near $x=0$ to derive asymptotic approximations in different regimes. 

%%%%%%%%%%%%%%%%%%%%%%%%%%%%%%%%%%%%%%%%%%%%%%%%%%%%%%%%%%%%%%
\begin{figure}[t!]
\includegraphics[width=1.0\columnwidth]{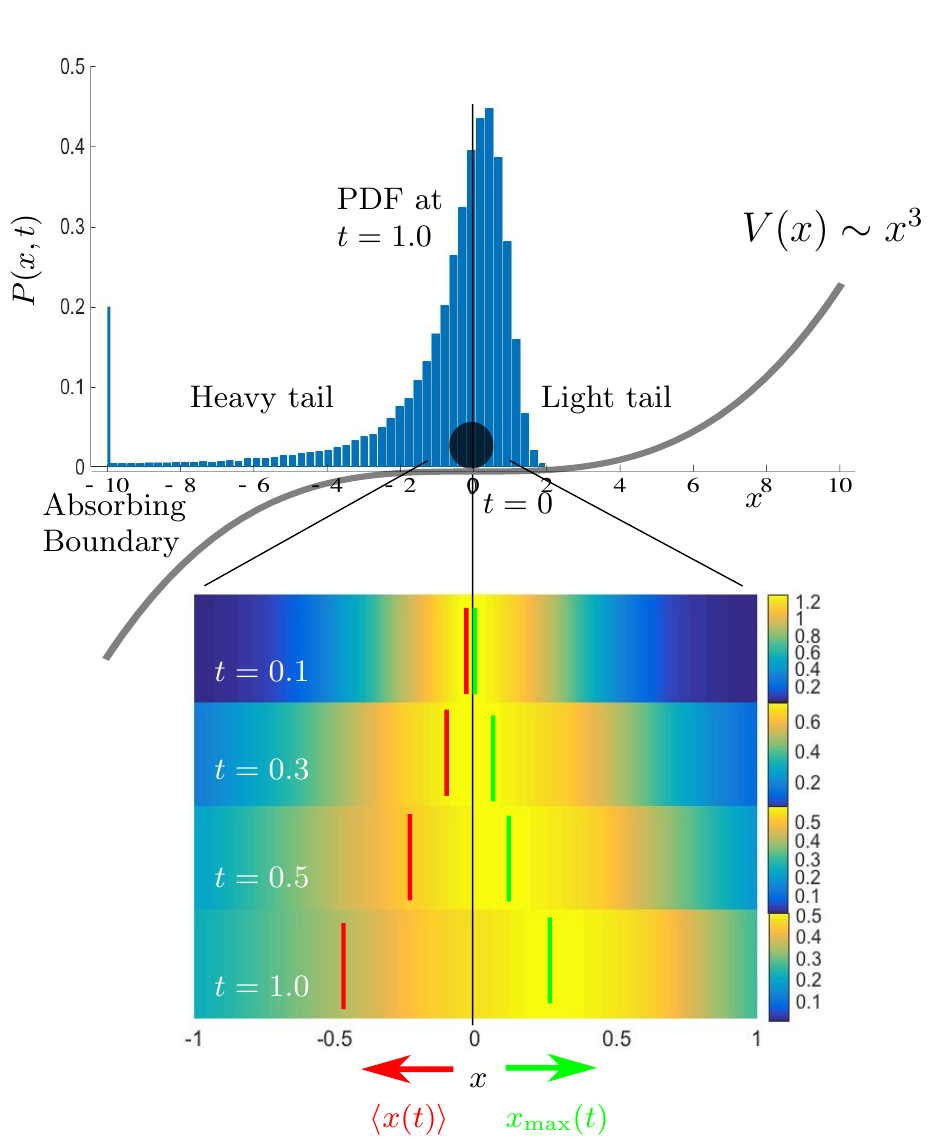}
\caption{\label{fig:illustration1} Difference between the regular local and the divergent global statistical descriptions of the Brownian motion in the unstable cubic potential. A particle is initially placed at the inflection point (black circle). In the global description (mean position) depicted by the red lines in density plots, the mean is quickly dragged towards $-\infty$ due to the  instability and hence it can be used for a very short time only. That behavior is also reflected in the presence of the heavy tail of $P(x,t)$ (upper panel). On the other hand, the maximum of $P(x,t)$ (green lines, the local description) moves atypically in the direction opposite to the acting force. The instability at negative $x$ does not invalidate the latter quantity even for long times. This gives the possibility to go beyond statistical moments in the local description of unstable motion. The higher moments vs.\ their local counterparts are discussed in Fig.~\ref{snr_temp}.
} 
\end{figure} 
%%%%%%%%%%%%%%%%%%%%%%%%%%%%%%%%%%%%%%%%%%%%%%%%%%%%%%%%%%%%%%

In the present work, we go beyond the aforementioned studies in the following ways. First, we argue that highly unstable dynamics~\eqref{Langevin} leads, already after a short time, to PDFs with heavy tails, which makes useless the description of $x(t)$ in terms of statistical moments. Instead, we propose to characterize the particle position by a directly measurable maximum of the PDF and use a curvature of the PDF at the maximum to characterize uncertainty. Second, we derive and discuss generic properties of the PDF including short-time dynamics, development of the heavy tail, and long-time properties, which turn out to be universal and described by the theory of {\em quasi-stationary distributions} \cite{PolletURL, bookQSD}. Quasi-stationary distributions emerge in stochastic processes conditioned on ``non-absorption''. Their study started by the seminal paper of Yaglom on the Galton-Watson branching process \cite{Yaglom1947}. Since then, the conditioned processes were successfully applied in mathematical biology \cite{Hastings2004}, epidemiology \cite{Nasell1995} and demographic studies \cite{Steinsaltz2004}, where the absorption corresponds to the extinction of a modeled population. The conditioning on non-absorption shifts focus on an ensemble of surviving individuals. In our context, the non-absorption roughly means that the particle remains on the potential plateau. The conditioning restricts our attention to trajectories which do not diverge up to a certain time.  In addition, we relate evolution of the PDF towards the quasi-stationary distribution to a mysterious creature known as Maxwell's demon \cite{bookLeffRex, Parrondo2015}.

All these main ideas are motivated and explained on physical grounds in the next section~\ref{sec:description}, which outlines the main results of our approach. All technical details concerning derivations and thorough discussions of particular points are comprised in the remaining sections \ref{sec:deterministic}-\ref{sec:quasistationary}.

%%%%%%%%%%%%%%%%%%%%%%%%%%%%%%%%%%%%%%%%%%%%%%%%%%%%%%%%%%%%%%%%%%%%%%%%%%%%%%%%%%%%%%%%%%%%%%%%%%%%%%%%%%%
%%%%%%%%%%%%%%%%%%%%%%%%%%%%%%%%%%%%%%%%%%%%%%%%%%%%%%%%%%%%%%%%%%%%%%%%%%%%%%%%%%%%%%%%%%%%%%%%%%%%%%%%%%%
\section{Pertinent description of Rapidly diverging trajectories}
\label{sec:description}
%%%%%%%%%%%%%%%%%%%%%%%%%%%%%%%%%%%%%%%%%%%%%%%%%%%%%%%%%%%%%%%%%%%%%%%%%%%%%%%%%%%%%%%%%%%%%%%%%%%%%%%%%%%
%%%%%%%%%%%%%%%%%%%%%%%%%%%%%%%%%%%%%%%%%%%%%%%%%%%%%%%%%%%%%%%%%%%%%%%%%%%%%%%%%%%%%%%%%%%%%%%%%%%%%%%%%%%

In the cubic potential~\eqref{potential}, the particle dynamics is considerably different for $|x(t)|< \left(3 k_{\rm B}T/k\right)^{1/3}$ and $|x(t)|> \left(3 k_{\rm B}T/k \right)^{1/3}$ \citep{ARPZRF2016}. Near the inflection point ($x=0$) on the potential plateau, the cubic potential is negligible compared to the thermal noise. Hence, when $|x(t)|< \left(3 k_{\rm B}T/k \right)^{1/3}$, the particle diffuses almost freely with only a weak drag to the left. On the other hand, the drag force rapidly increases in strength as $x(t)$ departs from the plateau. Actually, for $|x(t)|> \left( 3 k_{\rm B}T/k \right)^{1/3}$, the potential is so strong that the particle appearing at $x(t)<(3 k_{\rm B} T/k)^{1/3}$ reaches minus infinity {\em in a finite time} \citep{Radim1, ARPZRF2016}. On the other hand, the particle at $x(t)>(3 k_{\rm B} T/k)^{1/3}$ is dragged extremely quickly to the plateau. 

%%%%%%%%%%%%%%%%%%%%%%%%%%%%%%%%%%%%%%%%%%%%%%%%%%%%%%%%%%%%%%
\begin{figure}[t!]
\includegraphics[width=1.0\columnwidth]{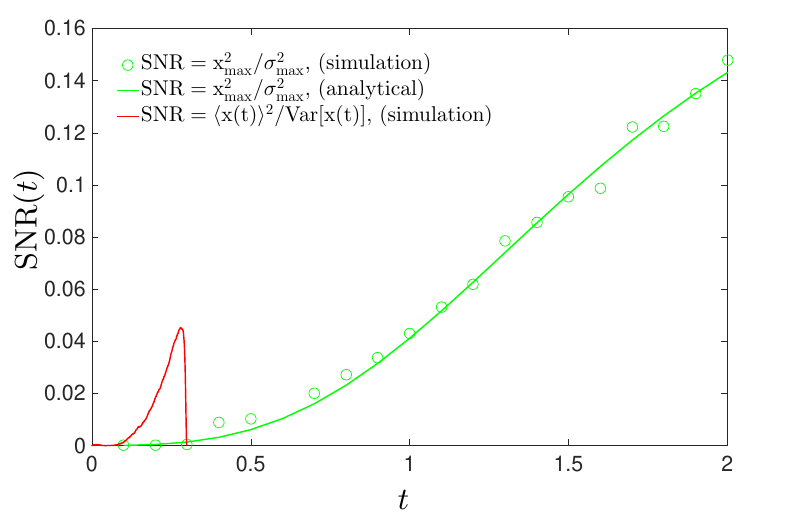}
\caption{\label{snr_temp}
Fast divergence of the global description using averages is demonstrated by a swift drop of the SNR (red line). The local description using the maximum of the PDF and the curvature at the maximum maintains its information value for all times (green line and $\circ$).   
In Monte Carlo simulations we have generated $4\times10^5$ trajectories with the time-step $\Delta t =0.002$ starting at the origin, $x(0)=0$, and diffusing with $D=0.1$ in the cubic potential with the stiffness $k=1$. Analytical result for the SNR~\eqref{snr} (green line) is derived in Sec.~\ref{sec:shorttimes}.  
The figure clearly demonstrates that with the local description of the system we can go beyond the statistical moments description which is reflected in growth of the SNR~\eqref{snr} (green line and~$\circ$).}
\end{figure}    
%%%%%%%%%%%%%%%%%%%%%%%%%%%%%%%%%%%%%%%%%%%%%%%%%%%%%%%%%%%%%%

The rapid divergence of trajectories implies unique features of the PDF $P(x,t)$. First of all, $P(x,t)$ develops a heavy tail for negative $x$ (as derived in Sec.~\ref{sec:deterministic}). This renders worthless the usual description of $x(t)$ in terms of moments $\left< x(t) \right> $, $\left< x^{2}(t) \right> $, \ldots, even at relatively short times. The higher the moment, in fact, the faster the divergence, which we illustrate in Fig.~\ref{snr_temp}, where the ratio 
$\left< x(t) \right>^2/ {\rm Var}[x(t)] $, $ {\rm Var}[x(t)] = \left< x^{2}(t) \right>- \left< x(t) \right>^{2}$,  
is plotted by the blue line. Because of the divergence, the ratio quickly drops to zero \cite{Radim1}. 
Assuming $\langle x(t)\rangle$ as an average useful signal from the unstable dynamics, this ratio can be interpreted as a signal-to-noise ratio (SNR). A drop of the SNR means that the signal in the position is negligible compared to the noise.   

It is therefore necessary to adopt a description of the present unstable system, which goes beyond the statistical moments. The main idea is to focus on the most probable particle position, i.e., on the position of maximum of $P(x,t)$, $x_{\rm max}(t)$, (instead of the mean value) and on the local curvature of $P(x,t)$, (instead of the variance). 
This approach has already been used to define local uncertainty for non-Gaussian distributions in quantum optics \cite{RadimPRA2013}. In the present model, this choice is experimentally motivated. It corresponds to a picture obtained from a detector linearly sensitive to larger density of particles (or trajectories) above some minimum threshold, as depicted in Fig.~\ref{fig:illustration1}. 

This measurement bears little information about diverging trajectories and provides a coherent picture of the most probable particle position near the instability. 
To quantify the relative fluctuations near the most probable position, we define the ``signal-to-noise'' ratio
\begin{equation}
\label{snr}
{\rm SNR}(t) = \frac{x_{\rm max}^2(t)}{\sigma_{\rm max}^2(t)},
\end{equation}
where we have introduced the normalized inverse curvature at the maximum  \cite{RadimPRA2013},
\begin{equation}
\label{sigmadef}
\sigma_{\rm max}^2(t) = \frac{P(x_{\rm max}(t),t)}{\left|\partial^{2}_{xx} P(x_{\rm max}(t),t)\right|}.
\end{equation} 
Note that for a Gaussian distribution the inverse curvature~\eqref{sigmadef} equals to the variance, $\sigma_{\rm max}^{2}(t)={\rm Var}[x(t)]$. The inverse curvature can be experimentally reached \cite{RadimPRA2013} following a  conditional version of the Central limit theorem \cite{CramerEisert2010}. We also note that an alternative regularized description based on quantiles (the median and quartiles) of the position distribution is possible. We leave a discussion of advantages and disadvantages of this possibility to a further study.  

SNR~\eqref{snr} specifies how well the most likely position can be observed in an experiment. It is a crucial parameter
for a possible experimental test of our results. As we discuss below, $x_{\rm max}$ is shifted to the right from $x=0$. This shift will be experimentally detectable only if the SNR is not negligible.
SNR~\eqref{snr} is shown in Fig.~\ref{snr_temp}. In contrast to the ratio of averages $\langle x(t) \rangle^{2} / \langle x^{2}(t) \rangle$, it shows no drop as time grows. In fact, the SNR~\eqref{snr} remains nonzero for any $t$, because both the maximum and the local curvature converges to a  positive value.  In contrast to this, the average particle position always moves in the direction of the force, cf.\ Fig.~\ref{fig:illustration1}. 

The second key feature of the PDF $P(x,t)$ induced by high instability of the potential \eqref{potential}, is that $P(x,t)$ is not normalized to one on the real line $x\in (-\infty,+\infty)$. The normalization 
\begin{equation}
\label{St}
S(t)=\int_{-\infty}^{\infty}dx P(x,t), 
\end{equation}
known as the survival probability \cite{bookRedner}, gives weight of trajectories that have not reached $x=-\infty$ by the time $t$. The survival probability decays with time exponentially when $D > 0$ (Sec.~\ref{sec:quasistationary}). Thus, in an ensemble of trajectories, the total weight of those wandering on the potential plateau decreases as individual trajectories are quickly dragged towards minus infinity. This phenomenon can be well understood in the analytically tractable case of $D=0$ discussed in Sec.~\ref{sec:deterministic}. Simultaneously, for $D=0$, the instability causes that $P(x,t)$ vanishes for $x>1/\kappa t$. 

The third intriguing feature of the present unstable system is that $P(x,t)$ quickly attains a universal spatial shape, $P(x,t)\sim  Q_{\rm st}(x) {\rm e}^{- \lambda_0 t}$, where $\lambda_0>0$ determines the decay rate of the unstable state. The normalized PDF $Q_{\rm st}(x)$ is the long-time limit of the ratio
\begin{equation}\label{eq6} 
Q(x,t) =\frac{P(x,t)}{S(t)}.
\end{equation} 
For any given $x$, the PDF $P(x,t)$ decays exponentially with time. Consequently, the survival probability $S(t)$,  Eq.~\eqref{St}, which is just the normalization of $P(x,t)$, also decays to zero. However, their ratio~\eqref{eq6} converges to the time-independent normalized distribution $Q_{\rm st}(x)$, which is  known as the {\em quasi-stationary distribution}  \cite{PolletURL, bookQSD}. The PDF $Q(x,t)$ is the conditional distribution of particles which do not reach $x=-\infty$ before time $t$. Its long-time limit $Q_{\rm st}(x)$ thus describes statistics of long-living (living = not diverging) trajectories. Note that $Q(x,t)$ and $P(x,t)$ are proportional and thus the maximum and the curvature of the both PDFs are the same.

Hence, in the long-time limit, the quasi-stationary distribution $Q_{\rm st}(x)$ provides analytical estimate of the local curvature of the generic PDF $P(x,t)$ around its maximum. Its position, $x_{\rm max}$, nontrivially depends on both the potential $V(x)$ and the temperature $T$. Interestingly, the curvature at the maximum of $Q_{\rm st}(x)$ ($1/\sigma_{\rm max}^{2}$, derived in Sec.~\ref{sec:quasistationary}) is determined by two qualitatively different factors,
\begin{equation} 
\label{InvCurvQS}
\frac{1}{\sigma_{\rm max}^{2}} = \frac{ V''(x_{\rm max})}{k_{\rm B} T} + \frac{\lambda_0}{D}.
\end{equation}
The first term on the right-hand side, $V''(x_{\rm max})/k_{\rm B} T$ is the (scaled) curvature of the potential. In our case it equals $V''(x_{\rm max})/k_{\rm B} T = 2k x_{\rm max}/k_{\rm B} T$. This first term alone determines curvature of any PDF of the functional form $p(V/k_{\rm B} T)$ (such as the Gibbs equilibrium distribution). 
The second term, $\lambda_0/D$, is always positive. Its magnitude is related to degree of instability of the system quantified by the decay rate $\lambda_0$. Thus, the quasi-stationary distribution is always narrower near its maximum than any PDF $p(V /k_{\rm B} T)$. The more unstable the system is (large decay rate $\lambda_0$), the narrower the distribution $Q_{\rm st}(x)$ becomes.

The analytical result~\eqref{InvCurvQS} has also two practical consequences. First, Eq.~\eqref{InvCurvQS} provides an independent scheme for measurement of the local curvature $1/\sigma_{\rm max}^{2}$. This is important since a direct inference of $1/\sigma_{\rm max}^{2}$ from the experimental data may depend on the fitting procedure used. Measuring curvature according Eq.~\eqref{InvCurvQS} avoids fitting of the PDF. Instead it uses easily-accessible first-passage properties, e.g.\ the survival probability, to determine the decay rate $\lambda_0$, which can be reliably measured  even for small samples of trajectories, see Ref.~\cite{SciRep2017}. Second, the result~\eqref{InvCurvQS} allows to extract scaling of the curvature with the intensity of thermal noise, $x_{\rm max}\sim (k_{\rm B} T)^{1/3}$, $\sigma^{2}_{\rm max} \sim (k_{\rm B} T)^{2/3}$, cf.\ Eqs.~\eqref{scalingX_Sigma}, that allows us to immediately find the SNR \eqref{snr} to be temperature-independent.

Last, but not least, the quasi-stationary distribution can be interpreted as a steady state PDF which we will explain in Sec.~\ref{sec:quasistationary}. Surprisingly, to accomplish this task we will need to introduce the feedback mechanism which we can interpret as the action of a Maxwell's demon.

%%%%%%%%%%%%%%%%%%%%%%%%%%%%%%%%%%%%%%%%%%%%%%%%%%%%%%%%%%%%%%%%%%%%%%%%%%%%%%%%%%%%%%%%%%%%%%%%%%%%%%%%%%%
%%%%%%%%%%%%%%%%%%%%%%%%%%%%%%%%%%%%%%%%%%%%%%%%%%%%%%%%%%%%%%%%%%%%%%%%%%%%%%%%%%%%%%%%%%%%%%%%%%%%%%%%%%%
\section{Instability yields heavy tails and decay of normalization of $P(x,t)$}
\label{sec:deterministic}
%%%%%%%%%%%%%%%%%%%%%%%%%%%%%%%%%%%%%%%%%%%%%%%%%%%%%%%%%%%%%%%%%%%%%%%%%%%%%%%%%%%%%%%%%%%%%%%%%%%%%%%%%%%
%%%%%%%%%%%%%%%%%%%%%%%%%%%%%%%%%%%%%%%%%%%%%%%%%%%%%%%%%%%%%%%%%%%%%%%%%%%%%%%%%%%%%%%%%%%%%%%%%%%%%%%%%%%

%%%%%%%%%%%%%%%%%%%%%%%%%%%%%%%%%%%%%%%%%%%
% FIG PDF D=0
%%%%%%%%%%%%%%%%%%%%%%%%%%%%%%%%%%%%%%%%%%%
\begin{figure}[t!]
\includegraphics[width=1.0\columnwidth]{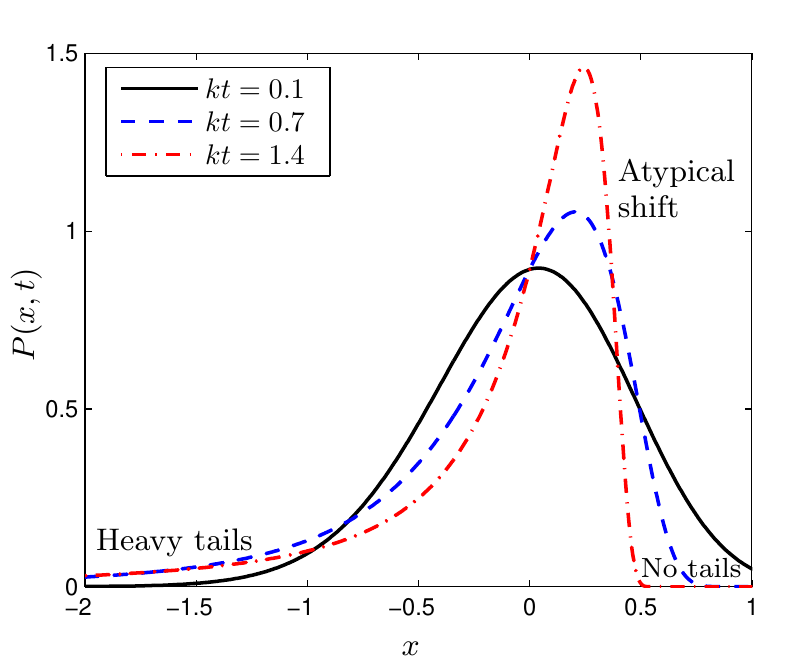}
\caption{\label{fig:PD0} PDF~\eqref{pdfD0} in three different times for $D=0$. The PDF~\eqref{pdfD0} develops a left heavy tail starting from the initial Gaussian distribution with the mean $x_0=0$ and the variance $\sigma^{2}_0=0.2$. For $x>1/\kappa t$, $P(x,t)$ is equal to zero (``No tails'' for $x>0$) due to a high speed of dynamics generated by the cubic potential. The maximum shifts in the opposite direction than the force acts and local uncertainty around the maximum decreases. Nonmonotonic behavior of the maximum, observed for longer times and different $x_0$, is further illustrated in Fig.~\ref{fig:maxD0}.}
\end{figure}
%%%%%%%%%%%%%%%%%%%%%%%%%%%%%%%%%%%%%%%%%%%

The simplified situation with negligible thermal noise ($D= 0$) is particularly useful, because it illustrates (i) the development of the heavy tail of $P(x,t)$ for negative $x$, (ii)  the vanishing of $P(x,t)$ for large $x$ in a finite time, (iii) an atypical shift of the PDF maximum and (iv) it elucidates properties of the survival probability~\eqref{St}.  PDF for the  present deterministic dynamics becomes non-trivial if we require a suitable initial distribution. We choose $P(x,0)$ to be Gaussian with the mean $x_0$ and the variance $\sigma_0^2$,
\begin{equation} 
\label{Gaussian}
P(x,0) =  \frac{{\rm e}^{-\frac{\left(x- x_0 \right)^2}{2\sigma_0^2} }}{\sqrt{2\pi \sigma_0^2}}.
\end{equation}
Then, at time $t$, $t>0$, we get the PDF \cite{Radim1} 
\begin{equation} 
\label{pdfD0}
P(x,t) = \theta\left(1/\kappa t-x\right) \frac{\exp \left\{-\frac{1}{2\sigma_0^2} \left(\frac{x}{1-x \kappa t}- x_0 \right)^2\right\}}{\sqrt{2\pi \sigma_0^2}  (1- x \kappa t)^2},
\end{equation}
where $\kappa=k/\gamma$ and $\theta(\bullet)$ stands for the Heaviside theta function. Derivation of Eq.~\eqref{pdfD0} can be found in Ref.~\cite{Radim1}, where fast divergence of averages $\left< x(t) \right>$, $\left< x^{2}(t) \right>$ was thoroughly discussed. For large negative $x$, the distribution decreases as $1/x^{2}$ and hence its moments do not exist. Fig.~\ref{fig:PD0} illustrates the gradual increase of the left tail with time.

Strong instability of the cubic potential manifests itself also in another feature of the PDF~\eqref{pdfD0}. The Heaviside theta function in Eq.~\eqref{pdfD0} implies that $P(x,t)$ vanishes when $x>1/\kappa t$ even though the initial Gaussian distribution~\eqref{Gaussian} has the infinite  support $x\in (-\infty,+\infty)$. Thus,  at the time $t$,  there are no trajectories on the right from $x=1/\kappa t$. The cubic potential is so strong that all trajectories with $x(0)>0$ quickly aggregate on the potential plateau on the right of $x = 0$. This happens in a finite time, regardless the initial position of the trajectory. In Fig. \ref{fig:PD0}, we denote the depopulated region as ``No tails'', in contrast to the heavy tail for $x\to - \infty$.

An analogous picture holds to the left of the inflection point. Any trajectory that starts on the negative half line is quickly dragged towards $x=-\infty$. This can be seen from a decrease of the survival  survival probability~\eqref{St} with time. The survival probability, which is the probability to find the particle on $x\in (-\infty,\infty)$ (norm of the PDF $P(x,t)$ in Eq.\ \eqref{pdfD0}) is given by 
\begin{equation}
\label{StD0}
 S(t) = \frac{1}{2}\left[ 1+ {\rm erf}\left( \frac{1+x_0 \kappa t}{\sqrt{2 \sigma_{0}^{2}} \kappa t} \right) \right],
\end{equation}
and its long-time limit reads
\begin{equation}
\label{StD0Lim}
\lim_{t \to \infty} S(t) = \frac{1}{2}\left[ 1 + {\rm erf}\left(\frac{x_0}{\sqrt{2 \sigma_{0}^{2}}} \right) \right]. 
\end{equation} 
When the initial particle distribution is the delta function at $x_0$, i.e.,  for $\sigma_{0}=0$, the right-hand side of Eq.~\eqref{StD0Lim} depends solely on $x_0$ and reduces to a unit step function at $x_0=0$. A nonzero width of the initial Gaussian PDF, $\sigma_0 >0$, broadens the step function because even for $x_0<0$, the nonzero $\sigma_0$ allows to generate an initial position on the right of the origin.

Moreover the decay of the survival probability $S(t)$ when $D>0$ turns out to be exponential as it will be discussed in  section~\ref{sec:quasistationary}. Differently to these asymptotic features, local dynamics of maximum and curvature of $P(x,t)$, discussed in the following section, do not depend on normalization of $P(x,t)$.

%%%%%%%%%%%%%%%%%%%%%%%%%%%%%%%%%%%%%%%%%%%%%%%%%%%%%%%%%%%%%%%%%%%%%%%%%%%%%%%%%%%%%%%%%%%%%%%%%%%%%%%%%%%
%%%%%%%%%%%%%%%%%%%%%%%%%%%%%%%%%%%%%%%%%%%%%%%%%%%%%%%%%%%%%%%%%%%%%%%%%%%%%%%%%%%%%%%%%%%%%%%%%%%%%%%%%%%
\section{Transient dynamics of maximum and curvature}
\label{sec:shorttimes}
%%%%%%%%%%%%%%%%%%%%%%%%%%%%%%%%%%%%%%%%%%%%%%%%%%%%%%%%%%%%%%%%%%%%%%%%%%%%%%%%%%%%%%%%%%%%%%%%%%%%%%%%%%%
%%%%%%%%%%%%%%%%%%%%%%%%%%%%%%%%%%%%%%%%%%%%%%%%%%%%%%%%%%%%%%%%%%%%%%%%%%%%%%%%%%%%%%%%%%%%%%%%%%%%%%%%%%%

%%%%%%%%%%%%%%%%%%%%%%%%%%%%%%%%%%%%%%%%%%%%%%%%%%%%%%%%%%%%%%%%%%%%%%%%%%%%%%%%%%%%%% 
\subsection{Diffusionless case $(D=0)$}
%%%%%%%%%%%%%%%%%%%%%%%%%%%%%%%%%%%%%%%%%%%%%%%%%%%%%%%%%%%%%%%%%%%%%%%%%%%%%%%%%%%%%% 

%%%%%%%%%%%%%%%%%%%%%%%%%%%%%%%%%%%%%%%%%%%
% FIG Xmax, Smax D=0
%%%%%%%%%%%%%%%%%%%%%%%%%%%%%%%%%%%%%%%%%%%
\begin{figure}[t!]
\includegraphics[width=1.0\columnwidth]{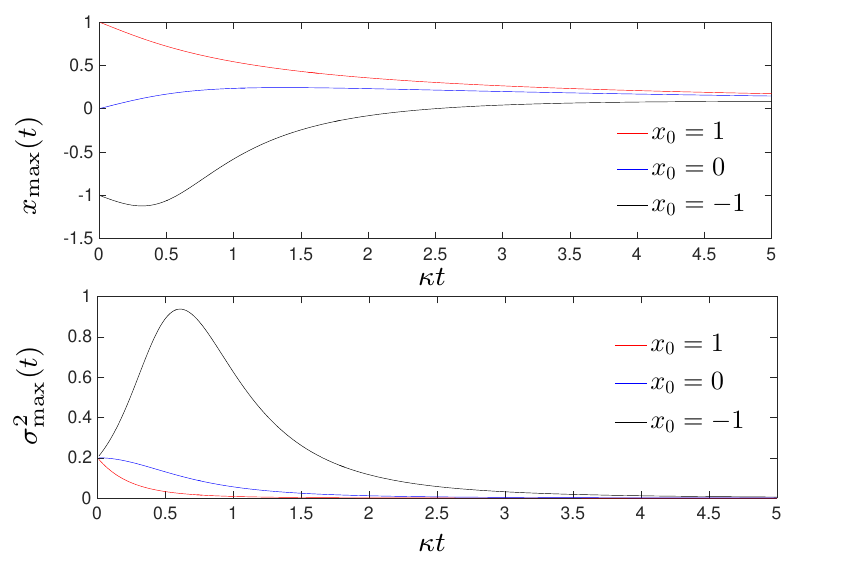}
\caption{\label{fig:maxD0} Evolution of the maximum and the inverse curvature of the PDF~\eqref{pdfD0} ($D=0$) for three initial Gaussian distributions with different mean $x_0$ and the same variance $\sigma_0=0.2$. 
For $x_0=1$, the maximum decreases towards $x=0$ and the inverse curvature quickly approaches zero. 
 For $x_0 = 0$, the maximum will first shift against the acting force and, after that, it will decrease back to $x = 0$. The curvature behaves similarly as in the previous case. 
 When $x_0=-1$, the maximum climbs above $x=0$ and converges back to $x=0$ at later times. The inverse curvature possesses a maximum. 
In all three cases, the long-time limit of $P(x,t)$ is the delta function at the origin with the weight given by the long-time survival probability~\eqref{StD0Lim}. 
The PDF for the case $x_0=0$ is shown in Fig.~\ref{fig:PD0}.   
}
\end{figure}
%%%%%%%%%%%%%%%%%%%%%%%%%%%%%%%%%%%%%%%%%%%

It is rather instructive to study the maximum and the curvature of the PDF~\eqref{pdfD0}. In contrast to the statistical moments, the two quantities describing the most probable particle position are not limited to short times. The position of the maximum of $P(x,t)$ is given by
\begin{equation}
\label{var}
x_{\rm max}(t)=\frac{1}{\kappa t}+\frac{1 + x_0 \kappa t - \sqrt{ (1+ x_0 \kappa t)^{2} +8 \sigma_0^2 (\kappa t)^2}}{4  \sigma_0^2 (\kappa t)^3}.
\end{equation}
The inverse curvature $\sigma_{\rm max}^2(t)$ is derived according to its definition~(\ref{sigmadef}). The result is, however, rather involved and hence we do not report it explicitly.

The behavior of both quantities, illustrated in Fig.~\ref{fig:maxD0}, should be understood based on the following consideration: A trajectory that starts from $x(0)$ follows the deterministic equation $x(t)=x(0)/(1+x(0) \kappa t)$. If the particle is initially located on the left from the inflection point $x=0$, it is quickly dragged towards $-\infty$. A particle located initially on the right of $x=0$, converges towards the origin as $x(t) \approx 1/\kappa t$. The trade-off between the two kinds of trajectories in the statistical ensemble determines all properties of $P(x,t)$. Surprisingly, this trade-off leads to a rich behavior of $x_{\rm max}(t)$ and $\sigma_{\rm max}^2(t)$, which strongly depends on the parameters of the initial distribution.

Further analytical insight for the case of non-vanishing $x_0$ can be gained for small times. For $t\to 0$ we have 
\begin{align}
\label{smalltX}
x_{\rm max}(t) & \approx x_0 + ( 2\sigma_0^2 - x_0^2 ) \kappa t, \\
\sigma_{\rm max}^2(t) & \approx \sigma_0^2 - 4 \sigma_0^2 x_0 \kappa t - 10 k^2 \sigma_0^4 t^2,
\end{align}
The inequality $0< x_0/\sqrt{2}<\sigma_0$ is a sufficient condition to observe the atypical shift of $x_{\rm max}$ against the acting force, $-V'(x)$. To observe the narrowing of $\sigma_{\rm max}^2$, it is then sufficient to have $x_0>0$. 
The equations justify qualitatively similar short-time decrease of $x_{\rm max}(t)$ for $x_0=\pm 1$ shown in Fig.~\ref{fig:maxD0} and also the initial increase of $\sigma_{\rm max}^2(t)$ for $x_0=-1$ and its decrease for $x_0=1$.
For $x_0=0$, $x_{\rm max}$ always evolves atypically and the inverse curvature in Eq.\ \eqref{smalltX} always decreases. 
The two characteristics also demonstrate an interesting nonlinear effect, namely the transformation of the initial variance (noise) into directed motion (notice the appearance of $\sigma_{0}^2$ in Eq.~\eqref{smalltX}), see Fig.~\ref{fig5} (inset). This effect is absent for the quadratic and the linear potential, where the corresponding Langevin equations are linear. 

In the long-time limit, the peak of the PDF $P(x,t)$ slowly sharpens and moves towards the origin from the right because all  trajectories, starting at $x(0)>0$, are sliding towards $x=0$. The tendency is  clearly visible in Fig.~\ref{fig:PD0}. Thus, $x_{\rm max}(t)\approx 1/\kappa t$ converges to zero and also $\sigma_{\rm max}^2(t) \sim (1/\kappa t)^{4} $ as the peak becomes sharper. In contrast to this, mean and variance are not defined for such a long period of time. 

Therefore the SNR~\eqref{snr} calculated for the PDF~\eqref{pdfD0} depends linearly on $\sigma_0^2$ in the short-time approximation, and it behaves qualitatively similar to its $D\neq 0$ counterpart depicted in Fig.\ \ref{snr_temp}. In the long-time limit this SNR grows as $t^2$ (which guarantees usefulness of the local description) as the span of the PDF on the positive half-line $x>0$ shrinks.

%%%%%%%%%%%%%%%%%%%%%%%%%%%%%%%%%%%%%%%%%%%%%%%%%%%%%%%%%%%%%%%%%%%%%%%%%%%%%%%%%%%%%% 
\subsection{Small diffusion case  $(D \ll 1)$}
%%%%%%%%%%%%%%%%%%%%%%%%%%%%%%%%%%%%%%%%%%%%%%%%%%%%%%%%%%%%%%%%%%%%%%%%%%%%%%%%%%%%%% 
 
%%%%%%%%%%%%%%%%%%%%%%%%%%%%%%%%%%%%%%%%%%% 
% FIG
\begin{figure}[t!]
\includegraphics[width=1.0\columnwidth]{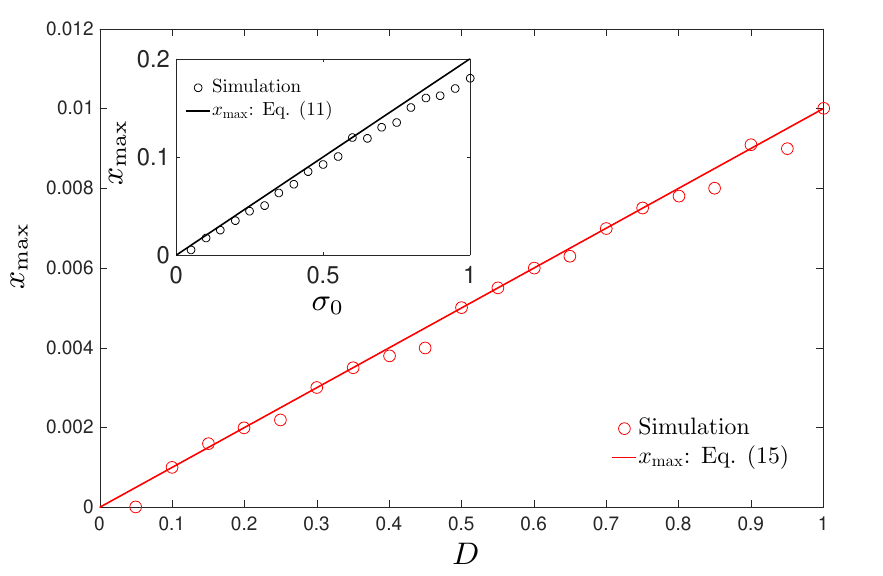}
\caption{\label{fig5} 
Atypical shift of the maximum at a given time induced by increasing temperature (i.e., $D$) or the initial variance $\sigma^{2}_0$ (inset). In simulations, we used $k=1$, $x_0=0$, the time-step $\Delta t = 0.002$, $\sigma_0=0$ (and $D=0$ for the inset) and $t=0.1$; $3\times10^5$ trajectories were generated.  The small $D$ approximation used to plot the red line provides a satisfactory result also for $D\approx 1$. The both plotted dependencies are predicted by two approximate equations, Eq.~\eqref{smalltX} (black line, inset) and Eq.~\eqref{maxTemp} (red line). Note that the SNR~\eqref{snr} grows linearly both with $D$ and $\sigma_0^2$.
} 
\end{figure} 
%%%%%%%%%%%%%%%%%%%%%%%%%%%%%%%%%%%%%%%%%%%%

\begin{figure}[t!]
\includegraphics[width=1.0\columnwidth]{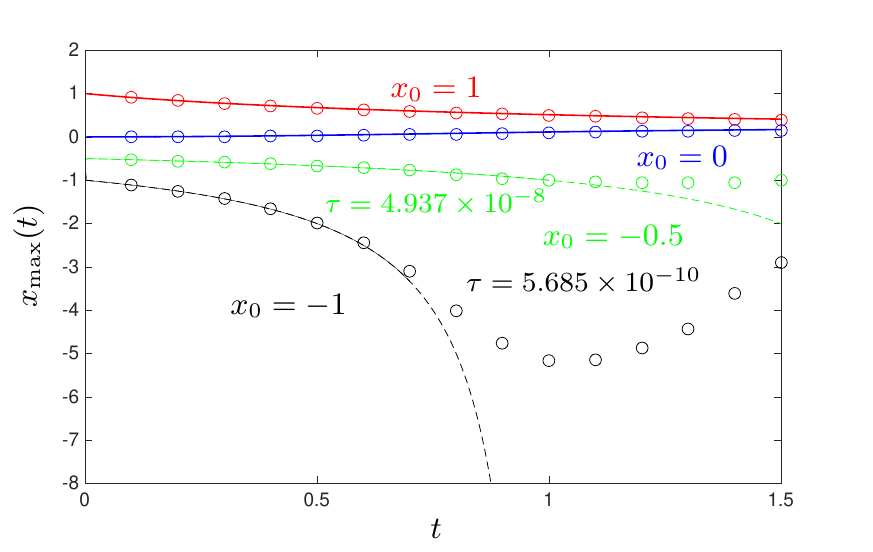}
\caption{\label{max_curv_t}  Evolution of the maximum in the small noise regime for different values of the initial particle position $x_0$. The depicted dependencies are qualitatively similar to their zero-noise ($D=0$) analogues from Fig.\ \ref{fig:maxD0}. 
When the initial position $x_0$ is zero or positive, $\tau =t$ (see main text). For negative $x_0$ values of $\tau$ are indicated in the graph by the corresponding color. In this graph we have used  $k=1$, $D=0.05$, and simulated $3\times 10^5$ trajectories with the time-step $\Delta t=0.002$.}
\end{figure}

For nonlinear potentials, the small noise expansion is not uniform in time \cite{bookGardiner}. Below, we present a trick how to extend the validity of the approximation, which is necessary when the particle starts on the left from the inflection point, $x_0<0$ (cf.\ Fig.~\ref{max_curv_t}). In the present section, we set variance of the initial distribution equal to zero, $\sigma_0^2=0$. Hence the only source of randomness is the (small) diffusion term in the Langevin equation~\eqref{Langevin}. 

The particle starts from $x_0$ on the potential plateau and its motion is initiated by a small thermal noise. It is reasonable to assume that after short initial period, the weak noise will play a negligible role as compared to the deterministic drift. The results from Ref.~\cite{Radim1} for the short-time averages read $\langle x(\tau) \rangle \approx x_0-\kappa  x_0^2 \tau- \kappa D\tau^2$ and ${\rm Var}[x(\tau)] \approx  2D\tau $, where $\tau$ will be treated as a small fitting parameter. In order to obtain the equation for the maximum in terms of the initial position and the time scale $\tau$, we substitute these moments into Eq.~\eqref{pdfD0}, $x_0 \to \langle x(\tau) \rangle$, $\sigma_0^{2} \to {\rm Var}[x(\tau)]$. After that we find the position of maximum of the PDF, 
\begin{equation}\label{max1Diff}
x_{\rm max}(t)\approx \frac{1+D\kappa^2t \tau_1 \tau - \sqrt{1+2D\kappa^2 t \tau_1 \tau+D^2\kappa^4 t^2\tau^4}}{8D\kappa^3t^3\tau},
\end{equation}
for $x_0=0$, where $\tau_1=8t-\tau$ (for $x_0 \neq 0$ the result is rather lengthy).

The approximation is compared with simulations in Fig.~\ref{max_curv_t}. Two qualitatively different regimes arise. The first occurs for $x_0 \geq 0$, where we are able to predict the dynamics for longer times and we do not need the fitting parameter $\tau$, i.e., $\tau  t$ in this case. The second type of dynamics with a different atypical effect occurs for $x_0<0$. Here, we fit $\tau$ to extend validity of the small noise approximation. Even so we are able to fit the data just before the turning point (the two lower curves in Fig.~\ref{max_curv_t}).

%%%%%%%%%%%%%%%%%%%%%%%%%%%%%%%%%%%%%%%%%%% 
% FIG
%%%%%%%%%%%%%%%%%%%%%%%%%%%%%%%%%%%%%%%%%%% 

To gain further insight into the role of the diffusion term, one can expand Eq.~\eqref{max1Diff} in series and notice that the maximum grows linearly with $D$, as shown in Fig.~\ref{fig5}, and quadratically with $t$,
\begin{equation}\label{maxTemp}
x_{\rm max}(t) \approx \kappa  Dt^2.
\end{equation} 
Even more interestingly, Eq.~\eqref{maxTemp} resembles the short-time limit of the first statistical moment, $\langle x(t) \rangle \approx - \kappa  D t^2$  \cite{Radim1}, but with the opposite sign.  
As can be seen directly from Eq.\ \eqref{maxTemp}, the bigger the diffusion parameter is the larger shift of the maximum is obtained. This behavior is shown in Fig.~\ref{fig5} for both the weak diffusion and the diffusionless case. The latter, presents dependence on the initial variance, instead of $D$. Contrary to Eq.\ \eqref{maxTemp}, the average would quickly diverge towards minus infinity, whereas the maximum, described by Eq.\ \eqref{max1Diff}, shifts in an opposite direction and converges to a finite value described in the next section. Focusing on the maximum (the most probable particle position) instead of the average $\langle x(t) \rangle$ thus allows us to avoid the singular properties of unstable dynamics.

The curvature \eqref{sigmadef}, calculated along similar lines as Eq.\ \eqref{max1Diff}, reads 
\begin{equation}\label{Curvature}
\sigma^2(t) \approx 2Dt-12D^2 \kappa^2t^4 ,
\end{equation}
where again a resemblance with the statistical moments can be seen in the first term, because we have  ${\rm Var}[x(t)] \approx  2Dt$ \cite{Radim1}.

Comparing Eqs.\ \eqref{max1Diff} and \eqref{Curvature}, one finds that the SNR~\eqref{snr} grows non-linearly in time as depicted in Fig.\ \ref{snr_temp} (green line) and for longer times it converges to a constant value.  Experimental observation of this and other atypical transient effects may require a fast detection of particle position during the transient period.

%%%%%%%%%%%%%%%%%%%%%%%%%%%%%%%%%%%%%%%%%%%%%%%%%%%%%%%%%%%%%%%%%%%%%%%%%%%%%%%%%%%%%%%%%%%%%%%%%%%%%%%%%%%
%%%%%%%%%%%%%%%%%%%%%%%%%%%%%%%%%%%%%%%%%%%%%%%%%%%%%%%%%%%%%%%%%%%%%%%%%%%%%%%%%%%%%%%%%%%%%%%%%%%%%%%%%%%
\section{Quasi-stationary distribution in the long-time limit}
\label{sec:quasistationary}
%%%%%%%%%%%%%%%%%%%%%%%%%%%%%%%%%%%%%%%%%%%%%%%%%%%%%%%%%%%%%%%%%%%%%%%%%%%%%%%%%%%%%%%%%%%%%%%%%%%%%%%%%%%
%%%%%%%%%%%%%%%%%%%%%%%%%%%%%%%%%%%%%%%%%%%%%%%%%%%%%%%%%%%%%%%%%%%%%%%%%%%%%%%%%%%%%%%%%%%%%%%%%%%%%%%%%%%

The discussed zero- and small noise approximations are not capable to capture properly long-time non-linear dynamics at the potential plateau (with non-negligible $D$). The reason is that even small noise affects significantly the long-time evolution, due to the high instability of the potential. Theoretical description, therefore, requires a different approach. 

%%%%%%%%%%%%%%%%%%%%%%%%%%%%%%%%%%%%%%%%%%%%%%%%%%%%%%%%%%%%%%%%%%%%%%%%%%%%%%%%%%%%%%%%%%%%%%%%%%%%%%%%%%%
\subsection{Definition and computation of \texorpdfstring{$Q_{\rm st}(x)$}{Qst}}
%%%%%%%%%%%%%%%%%%%%%%%%%%%%%%%%%%%%%%%%%%%%%%%%%%%%%%%%%%%%%%%%%%%%%%%%%%%%%%%%%%%%%%%%%%%%%%%%%%%%%%%%%%%

The cubic potential is highly unstable and hence one can hardly expect any nontrivial long-time behavior for the PDF $P(x,t)$. However, after a relatively short time the PDF $P(x,t)$ attains a universal shape determined by the function $Q_{\rm st}(x)$,
which is multiplied by a simple exponential decay in time, $P(x,t) \sim Q_{\rm st}(x) {\rm e}^{-\lambda_0 t}$. The normalized function $Q_{\rm st}(x)$, known as a quasi-stationary distribution \cite{PolletURL, bookQSD}, is independent of time and initial conditions. It is determined solely by the form of the potential. From the practical point of view, the quasi-stationary distribution can be used to characterize the unstable systems when moments fail and transients are too fast.  

In Monte Carlo simulations of individual trajectories, the quasi-stationary distribution is nothing but the normalized PDF of particles that are still on a finite $x$ at the time $t$, $t>\lambda_0$.
 Hence it should be understood as the long-time limit
\begin{equation}
Q_{\rm st}(x) = \lim_{t\to\infty} Q(x,t),
\end{equation} 
of the PDF conditioned on survival, cf.\ \eqref{eq6},
\begin{equation}
\label{Qdef}
Q(x,t) =\frac{P(x,t)}{S(t)},
\end{equation}
where both the nominator and the denominator (the norm of $P(x,t)$, cf.\ Eq.~\eqref{St}) tend to zero. The ratio, however, converges towards a finite value for any $x$. The function $Q(x,t)$ is the PDF of surviving trajectories (e.g.\ wandering on the potential plateau), which we described by the local measures in the previous sections.

%%%%%%%%%%%%%%%%%%%%%%%%%%%%%%%%%%%%%%%%%%% 
% FIG 
\begin{figure}[t!]
\includegraphics[width=1.0\columnwidth]{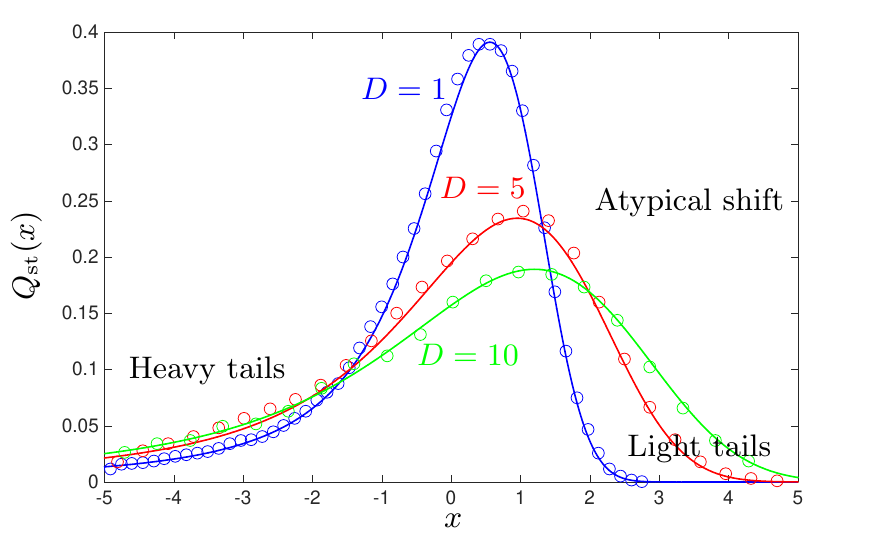}
\caption{\label{fig:Qst} Broadening of the quasi-stationary PDF~\eqref{eq:Qst_eigen} with increasing temperature. Solid lines depict numerical solutions of the eigenvalue problem~\eqref{Leqigenvalue} by the method explained in the appendix. Circles stand for outcomes of Monte Carlo simulations using $3\times10^5$ trajectories. In simulations, the particle starts from the inflection point and evolves (if not absorbed) for $t=7$ with the time-step $\Delta t =0.002$.  The stiffness $k$ of the cubic potential is set to one. Interesting effects due the instability observed in the transient dynamics have their analogues reflected  in the shape of the quasi-stationary PDF. Quasi-stationary PDF therefore can be used to describe unstable systems when averages diverge and transients are fast.   
}
\end{figure}  
%%%%%%%%%%%%%%%%%%%%%%%%%%%%%%%%%%%%%%%%%%% 

To derive equation for $Q_{\rm st}(x)$ we start from the Fokker-Planck equation corresponding to Eq.~\eqref{Langevin}. 
\begin{equation}
\label{FP}
\frac{\partial }{\partial t} P(x,t)= \mathcal{L} P(x,t),
\end{equation}
with the Fokker-Planck operator given by \cite{bookGardiner} 
\begin{equation}
\label{FPoperator}
\mathcal{L}= D \frac{\partial^{2}}{\partial x^{2}} + \frac{1}{\gamma} \frac{\partial}{\partial x} V'(x),
\end{equation} 
where $V'(x)$ stands for derivative of $V(x)$. 
We now introduce the Ansatz $P(x,t) \sim Q_{\rm st}(x) {\rm e}^{-\lambda_0 t}$
 into Eq.~\eqref{FP} together with the exponentially decaying survival probability~\eqref{St} and after some algebra we obtain that $Q_{\rm st}(x)$ is given  by 
 \begin{equation} 
Q_{\rm st}(x) = \frac{\psi_0(x)}{\int_{-\infty}^{\infty}dx\,\psi_0(x)},
\label{eq:Qst_eigen}
\end{equation}
where $\psi_0(x)$, the normalized eigenvector of $\mathcal{L}$ corresponding to its largest eigenvalue $-\lambda_0$, 
\begin{equation} 
\label{Leqigenvalue}
\mathcal{L} \psi_0(x) = -\lambda_0 \psi_0(x).
\end{equation}
For a rigorous proof we refer to \cite{bookQSD}. 
Quasi-stationary distribution, $Q_{\rm st}(x)$, is shown in Fig.\ \ref{fig:Qst}, for three different temperatures.

The eigenvalue problem~\eqref{Leqigenvalue} should be supplemented by boundary conditions. Interestingly enough, for the cubic potential, natural boundary conditions yield PDF which vanishes for $|x|\to \infty$, but the probability current does not vanish in the limit $x\to -\infty$. Thus, we can approximate the singular point $x=-\infty$ by placing an {\em absorbing boundary} \cite{bookRedner} at a finite position $x=a$, $a<0$. The absorbing boundary is nothing but a trap which captures (absorbs) the particle when it hits $x=a$ for the first time. In Fig.~\ref{fig:illustration1}, the boundary is at $x=-10$. The weight of absorbed trajectories increases with time and eventually tends to one. 

The regularization is convenient for numerical solution of \eqref{Leqigenvalue}, see the appendix, and is natural in Monte Carlo simulations. As long as $a \ll - (3 k_{\rm B}T/k )^{1/3}$ is satisfied, this cut-off will not affect properties of the slow stochastic motion on the plateau of the potential \eqref{potential}. Consequently, we can require $\psi_0(x)$ to satisfy the absorbing boundary condition $\psi_0(a)=0$ and the natural boundary condition at $x= \infty$.

Finally notice that $P(x,t) \sim Q_{\rm st}(x) {\rm e}^{-\lambda_0 t}$ determines just the main asymptotics of $P(x,t)$, i.e., the only significant term in the eigenvector expansion when $t\to \infty$. A time-dependent correction which describes relaxation towards the quasi-stationary distribution $Q_{\rm st}(x)$ decays exponentially fast, as ${\rm e}^{(\lambda_0 - \lambda_1)t}$. This is why in simulations $Q_{\rm st}(x)$  is readily observable for relatively short times. For counterexamples, where $Q(x,t)$ does not converge to a time-independent limit we refer, e.g.,\ to works~\cite{TracerPRE2014, LogarithmicJCP2015}.

%%%%%%%%%%%%%%%%%%%%%%%%%%%%%%%%%%%%%%%%%%%%%%%%%%%%%%%%%%%%%%%%%%%%%%%%%%%%%%%%
\subsection{\texorpdfstring{$Q_{\rm st}(x)$}{Qst} as a steady-state distribution and  Maxwell's demon}
\label{sec:demon}
%%%%%%%%%%%%%%%%%%%%%%%%%%%%%%%%%%%%%%%%%%%%%%%%%%%%%%%%%%%%%%%%%%%%%%%%%%%%%%%%

Usually, the term ``steady state" is related to the stationary long-time system state with time-independent currents \cite{BlytheEvans2007,Tailleur2008}. In particular, the Gibbs canonical equilibrium is an example of the isothermal steady state where all currents vanish. In more general nonequilibrium steady states the currents (in our case a probability current) converge to nonzero values which are closely related to local properties of the steady-state PDF. 
At a first glance, the quasi-stationary distribution $Q_{\rm st}(x)$ is not related to such scenario,  because there is no nontrivial long-time state in the unstable cubic potential (the particle, once released, reaches $x=-\infty$ in a relatively short time). The quasi-stationary PDF results from the limit of the ratio~\eqref{Qdef} of two vanishing terms and not as the result of balance of probability currents. 

The direct meaning of $Q_{\rm st}(x)$, according to its definition~\eqref{Qdef}, is that $Q_{\rm st}(x)$ stands for the PDF of a particle which survives (or, equivalently, stays on the potential plateau)  for a long time. However, it is rather the following steady-state interpretation which deepen our intuitive understanding of the model behavior and brings us straight to results for the maximum and curvature of $Q_{\rm st}(x)$. It can also inspire experimental method capable to reach quasi-stationary PDF using external control of the Brownian motion. To obtain the steady-state interpretation of the quasi-stationary distribution we first notice that the Fokker-Planck equation for $Q(x,t)$ reads
\begin{equation}
\label{FPcond}
\frac{\partial }{\partial t} Q(x,t) = \mathcal{L}Q(x,t) -  J_{Q}(a,t)  Q(x,t). 
\end{equation}
Eq.~\eqref{FPcond} follows from the Fokker-Planck equation~\eqref{FP} after inserting $P(x,t)=Q(x,t)S(t)$ into Eq.~\eqref{FP} and dividing the resulting equation by $S(t)$ (see Appendix~\ref{sec:appB} for more details). 

Above, $-J_{Q}(a,t)$ is the conditional probability current \cite{RiskenBook} into the absorbing boundary, 
\begin{equation}
\label{CondCurrent}
J_{Q}(x,t)= - \left( D \frac{\partial }{\partial x} + \frac{k}{\gamma} x^{2} \right)Q(x,t). 
\end{equation} 
The probability current $J_{Q}(a,t)$ is negative due to the sign convention (the current is positive when probability flows to the right) hence, the second term on the right-hand side of Eq.~\eqref{FPcond} represents the positive source of the probability. It ensures that the normalization of $Q(x,t)$ remains constant in time, in contrast to the Fokker-Planck equation~\eqref{FP} for the generic PDF $P(x,t)$, where such source term is missing and hence $P(x,t)$ is not normalized. The integral of this second term is exactly equal to the probability flow to the absorbing boundary.

%%%%%%%%%%%%%%%%%%%%%%%%%%%%%%%%%%%%%%%%%%%%%%%%%%%%%%%%%%%%%%%%%%%%%%%%%%%%%%%%
\begin{figure}[t!]
\includegraphics[width=1.0\columnwidth]{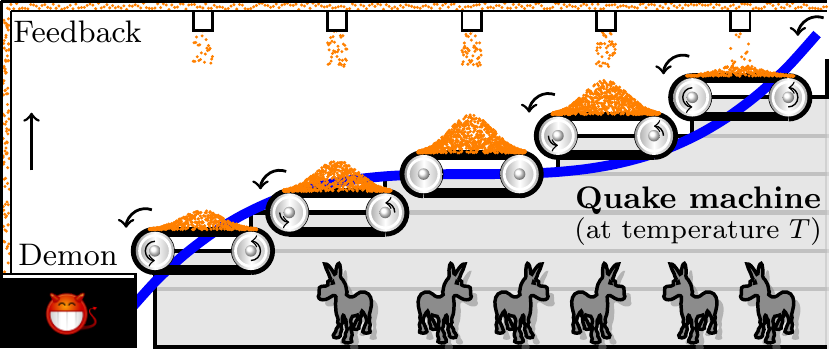}
\caption{Sketch of the steady-state factory producing $Q_{\rm st}(x)$ in the unstable cubic potential, where a standard stationary (equilibrium) distribution does not exist \cite{DonkeysURL}. Instead of tracking a single-particle trajectory as in previous figures, here we turn to the following experiment with many particles (sand). The demon (the measurement-feedback mechanism) collects the sand which leaves the system at its left boundary and returns it back according to Eq.~(\ref{FPcond}) such that the long-time distribution of the sand on the belts is given by $Q_{\rm st}(x)$. Detailed description of the machine stemming from  Eq.~(\ref{FPcond}) is presented in Sec.~\ref{sec:demon}.}
\label{fig:QuakeMachine}
\end{figure}
%%%%%%%%%%%%%%%%%%%%%%%%%%%%%%%%%%%%%%%%%%%%%%%%%%%%%%%%%%%%%%%%%%%%%%%%%%%%%%%%

The physical interpretation of Eq.~\eqref{FPcond} requires to describe a complex measurement and feedback mechanism restoring the normalization of $Q(x,t)$. The mechanism uses an ensemble of the particles, rather than just single particle, which we explain in the following. It is depicted using a cartoon style in Fig.~\ref{fig:QuakeMachine}. 
In the cartoon, the diffusing particles are represented by orange sand grains. Three basic ingredients, which drive the particles according to Eq.~\eqref{FPcond} and thus also the sand in the cartoon are: (i) The cubic potential approximated by five conveyor belts. The velocities of the belts are proportional to the gradient of the cubic potential at their positions (arrows on the rotating wheels). The belts outside the plateau of the potential systematically transport the sand from right to left, the belt at the plateau just collects the sand. (ii) The thermal noise represented by donkeys who randomly shake the conveyor holding structure (the Quake machine) as they stomp on its floor. At $T=0$ the donkeys are unflappable, they do not stomp and the shaking (the thermal motion) stops. Nonzero temperature corresponds to nervous donkeys, they stomp vigorously on the floor and the whole structure vibrates. The noise (vibrations), thus, affects globally the sand dynamics, but leaves intact the demon and feedback mechanism. Shaking causes sand grains to jump randomly from one belt to another, both to the left and to the right.

Formally, the two ingredients (i) and (ii) are included in the Fokker-Planck operator $\mathcal{L}$ \eqref{FPoperator}. The last part (iii) of the dynamics described by Eq.~\eqref{FPcond}, i.e.\ the absorbing boundary and the source term $-J_Q(a,t)Q(x,t)$, are depicted  by a black box with a Maxwell demon on the left from the conveyor belts. The demon acts both as a sink and as a source of the sand, namely it continuously monitors the number of sand grains on individual belts, accepts the sand which falls into the absorbing boundary from the leftmost conveyor belt, and instantaneously redistributes the accepted sand back to the belts. 
For the redistribution, the demon utilizes measured information about the instantaneous distribution of sand on all the belts. The demon is therefore continuously watching the whole factory.  The portions of sand which are delivered to individual belts are determined proportionally to the amount of sand presented on the belts at the time of redistribution. For example, the belt containing $10\%$ of all sand at the time of redistribution is refilled by $10\%$ of the redistributed sand at that time. This rule is a direct interpretation of the source term $-J_Q(a,t)Q(x,t)$ in Eq.~\eqref{FPcond}.

The total amount of sand in the system is fixed similarly as the norm of the PDF $Q(x,t)$ governed by Eq.~\eqref{FPcond}. After a relatively short time (determined by the inverse gap, $1/(\lambda_0-\lambda_1)$, between the two largest eigenvalues of the Fokker-Planck operator~\eqref{FPoperator}), the time-independent steady-state distribution of sand on the belts is established by balancing the sand (probability) currents caused by the three agents (i)-(iii) described above. The sand distribution, then, corresponds to the quasi-stationary PDF $Q_{\rm st}(x)$, for which the left-hand side of Eq.~\eqref{FPcond} vanishes. 

Comparing the resulting stationary Fokker-Planck equation with Eq.~\eqref{Leqigenvalue}, we get a noteworthy interpretation of the eigenvalue $-\lambda_0$. This inverse relaxation time is just the stationary conditional probability current into the absorbing boundary, 
\begin{equation}
\lambda_0 =  - J_{Q_{\rm st}}(a),
\label{LambdaJ}
\end{equation}
where $J_{Q_{\rm st}}(a) = \lim_{t\to\infty}J_Q(a,t)$. In other words, $\lambda_0$ measures the amount of sand per unit time which falls from the leftmost belt into the box (in the steady state).

Last, but not least, note that the above interpretation of Eq.~\eqref{FPcond} closely resembles stochastic processes with resetting, where particles are instantaneously returned to a certain position or region in space following a given protocol \cite{Manrubia1999, Evans2011,Kusmierz2014,Gupta2014,Murugan2014, Meylahn2015,Hartich2015,Eule2016,Pal2016,Fuchs2016, Pal2017, Montero2017}. This suggests that results found for systems with reseting can be readily used both in our model and in all similar scenarios, where one consider a probability density of surviving particles. Here we will evaluate the entropy flux extracted from the system by the Maxwell demon in order to sustain the quasi-stationary PDF $Q_{\rm st}(x)$.

We define the entropy of a surviving particle at time $t$ as $\mathcal{S}(t) = - k_{\rm B}\int_{-\infty}^{\infty}dx Q(x,t) \log Q(x,t)$. Taking the derivative with respect to time gives the entropy production $\dot{\mathcal{S}}(t) = - k_{\rm B}\int_{-\infty}^{\infty}dx\, \partial Q(x,t)/\partial t \log Q(x,t)$. Substituting for $\partial Q(x,t)/\partial t$ from Eq.~(\ref{FPcond}) into the last formula leads to the expression
\begin{equation}
\dot{\mathcal{S}}(t) = \dot{\mathcal{S}}_{\rm diff}(t) - \dot{\mathcal{S}}_{\rm Md}(t),
\label{eq:entropy}
\end{equation}
where
$\dot{\mathcal{S}}_{\rm diff}(t) = - k_{\rm B} \int_{-\infty}^{\infty}dx [\mathcal{L}Q(x,t)] \log Q(x,t)$, and $\dot{\mathcal{S}}_{\rm Md}(t) = -J_{Q}(a,t) \mathcal{S}(t)$.
The term $\dot{\mathcal{S}}_{\rm diff}(t)$ amounts for entropy increase due to diffusion in the cubic potential. The term $\dot{\mathcal{S}}_{\rm Md}(t)$ is the entropy flux out of the system due to the demon pushing the system towards the quasi-stationary PDF. After the system relaxes to the quasi-stationary state, i.e., for $Q(x,t) = Q_{\rm st}(x)$, the left-hand side of Eq.~(\ref{eq:entropy}) vanishes and thus the balance of the two entropy productions holds, $\dot{\mathcal{S}}_{\rm diff} =  \dot{\mathcal{S}}_{\rm Md}$. 
The amount of entropy the demon takes out of the system per unit time in order to sustain the non-equilibrium quasi-stationary state is thus proportional to the stationary entropy of the system and the stationary probability flux out of the system:
\begin{equation}
\label{stacS}
\dot{\mathcal{S}}_{\rm Md} = - J_{Q_{\rm st}}(a) \mathcal{S}_{{\rm st}} = \lambda_0  \mathcal{S}_{{\rm st}},
\end{equation}
where $
\mathcal{S}_{{\rm st}} = - k_{\rm B}\int_{-\infty}^{\infty}dx\, Q_{\rm st}(x) \log Q_{\rm st}(x)$. 
Eq.~\eqref{stacS} illustrates another important role of the relaxation rate $\lambda_0$.

%%%%%%%%%%%%%%%%%%%%%%%%%%%%%%%%%%%%%%%%%%%%%%%%%%%%%%%%%%%%%%%%%%%%%%%%%%%%%%%%%%%%%%%%%%%%%%%%%%%%%%%%%%%
%%%%%%%%%%%%%%%%%%%%%%%%%%%%%%%%%%%%%%%%%%%%%%%%%%%%%%%%%%%%%%%%%%%%%%%%%%%%%%%%%%%%%%%%%%%%%%%%%%%%%%%%%%%
\section{Quasi-stationary values of maximum and curvature}
%%%%%%%%%%%%%%%%%%%%%%%%%%%%%%%%%%%%%%%%%%%%%%%%%%%%%%%%%%%%%%%%%%%%%%%%%%%%%%%%%%%%%%%%%%%%%%%%%%%%%%%%%%%
%%%%%%%%%%%%%%%%%%%%%%%%%%%%%%%%%%%%%%%%%%%%%%%%%%%%%%%%%%%%%%%%%%%%%%%%%%%%%%%%%%%%%%%%%%%%%%%%%%%%%%%%%%%

%%%%%%%%%%%%%%%%%%%%%%%%%%%%%%%%%%%%%%%%%%% 
% FIG
\begin{figure}[t!]
\includegraphics[width=1.0\columnwidth]{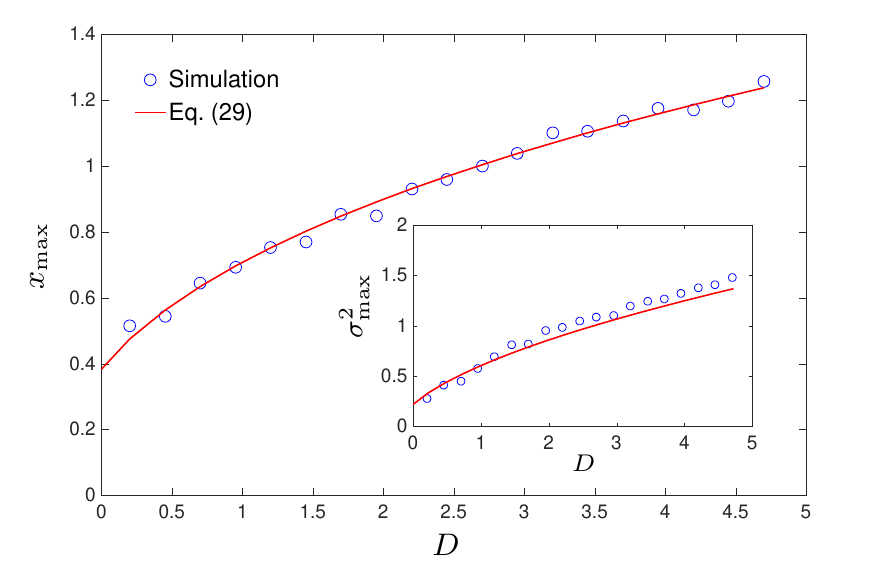}
\caption{\label{fig:XmaxSt} The maximum and the inverse curvature (inset) of $Q_{\rm st}(x)$ as functions of temperature. The both quantities follow exactly the scaling~\eqref{scalingX_Sigma}. 
Solid lines depict numerical solutions of the eigenvalue problem~\eqref{Leqigenvalue}, cf.\ the appendix. Circles represent simulated data using $3\times10^5$ trajectories. In simulations, the particle starts from the inflection point and evolves (if not absorbed) for $t=7$ with the time-step $\Delta t =0.002$.  The stiffness $k$ of the cubic potential is set to one.
}
\end{figure}
%%%%%%%%%%%%%%%%%%%%%%%%%%%%%%%%%%%%%%%%%%% 

The maximum of $Q_{\rm st}(x)$ and that of the generic PDF $P(x,t)$ (see Fig.~\ref{fig:illustration1}) coincide after a relatively short time. Its exact position $x_{max}$, however, depends on the potential and temperature in a non-trivial way.  On the other hand, for the curvature at the maximum, $1/\sigma^{2}_{\rm max}=|Q_{\rm st}''(x_{\rm max})|/Q_{\rm st}(x_{\rm max})$, we obtain from the stationary version  of the Fokker-Planck equation~\eqref{FPcond},
\begin{equation} 
\frac{1}{\sigma_{\rm max}^{2}} = \frac{ V''(x_{\rm max})}{k_{\rm B} T} + \frac{\lambda_0}{D}.
\label{Sigma2}
\end{equation}
Namely, we obtain Eq.~\eqref{Sigma2} from Eq.~\eqref{FPcond} after setting $\partial Q_{\rm st}/\partial t = 0$, $Q_{\rm st}'(x_{\rm max})=0$, and using Eq.~\eqref{LambdaJ} in the second term on the right-hand side. 

The above equation provides us an indirect and independent way how to determine the local width of the generic PDF $P(x,t)$. It is enough to find position of the maximum $x_{\rm max}$ and measure the decay rate $\lambda_0$. The latter measurement would be analogous to our recent experiments \cite{SciRep2017}, since it is enough to determine the decay rate of the survival probability, $S(t) \approx s_0 {\rm e}^{- \lambda_0 t}$. Moreover, from results of \cite{SciRep2017} it follows that the survival probability is easily measurable in the highly unstable potentials. Such independent measurement is needed because the direct determination of the local width is sensitive to the procedure used for fitting the PDF from experimental data 

The result~\eqref{Sigma2} is notable also for its physical content. Interestingly, the more unstable the system is (large $\lambda_0$), the narrower is the PDF around the maximum. Instability of the system can be controlled both by the strength of the thermal noise $D$ and by the amplitude $k$ of the cubic potential. Strong cubic potentials (larger $k$ for a given $D$) are more unstable and the plateau region bounded approximately to the interval $[-(3 k_{\rm B}T/k )^{1/3},(3 k_{\rm B}T/k )^{1/3}]$ is small in this case. The local width of the PDF for more unstable potentials decreases.

Temperature-dependence of the quasi-stationary PDF can be understood from scaling arguments \citep{ARPZRF2016, Dean1}. When the absorbing boundary is far from the origin, $a\ll -(3 k_{\rm B}T/k )^{1/3}$, there remain only two length scales in the problem: the width of the plateau and the thermal length dictated by $D$. The quasi-stationary PDF should depend on their ratio, and hence any length in the problem scales as $(k_{\rm B} T/ k )^{1/3}$. This is exactly what we observe for the maximum. Similar relation also holds for the local width of the PDF, 
\begin{equation}
\label{scalingX_Sigma}
x_{\rm max}\sim D^{1/3}, 
\qquad \sigma^{2}_{\rm max} \sim D^{2/3}.
\end{equation}
The maximum of the quasi-stationary PDF climbs up to higher values of the potential for higher temperatures and the local width at maximum increases. Both dependencies are demonstrated in Fig.~\ref{fig:XmaxSt}. The scaling implies that the SNR \eqref{snr} remains temperature independent.

%%%%%%%%%%%%%%%%%%%%%%%%%%%%%%%%%%%%%%%%%%%%%%%%%%%%%%%%%%%%%%%%%%%%%%%%%%%%%%%%%%%%%%%%%%%%%%%%%%%%%%%%%%%
%%%%%%%%%%%%%%%%%%%%%%%%%%%%%%%%%%%%%%%%%%%%%%%%%%%%%%%%%%%%%%%%%%%%%%%%%%%%%%%%%%%%%%%%%%%%%%%%%%%%%%%%%%%
\section{Concluding remarks and experimental perspectives}
%%%%%%%%%%%%%%%%%%%%%%%%%%%%%%%%%%%%%%%%%%%%%%%%%%%%%%%%%%%%%%%%%%%%%%%%%%%%%%%%%%%%%%%%%%%%%%%%%%%%%%%%%%%
%%%%%%%%%%%%%%%%%%%%%%%%%%%%%%%%%%%%%%%%%%%%%%%%%%%%%%%%%%%%%%%%%%%%%%%%%%%%%%%%%%%%%%%%%%%%%%%%%%%%%%%%%%%

Unstable systems are important for their potential applications. However, their description and characterization is challenging even in simplest cases. In the present work, we have developed a statistical description of position of a Brownian particle diffusing in the cubic potential. The task was complicated due to a high instability and nonlinearity of the model. As a consequence, the PDF of the particle position develops a heavy tail and its moments cease to exist. 
In this work we have proposed an appropriate, experimentally accessible description focusing on the most probable position of the particle (position of maximum of the PDF) and on a local curvature of the PDF at the maximum (instead of the variance). In contrast to the standard approach, which uses the moments, the two quantities are well defined even though the lifetime of any initial state is very short. We have described both the short-time (Sec.~\ref{sec:shorttimes}) and the long-time (Sec.~\ref{sec:quasistationary}) properties of the two quantities, both from an analytical and numerical perspective, with an emphasis on their time- and temperature- dependencies. Our results are general for unstable potentials with an inflection point and should be easily observable directly using position detectors in experiments similar to that reported in Ref.~\citep{SciRep2017}. 

In particular, the most probable position shows a peculiar behavior. The maximum of PDF can move opposite to the acting force both as the function of time and temperature (Figs.~\ref{fig:maxD0},~\ref{max_curv_t}, and~\ref{fig:XmaxSt}). 
The curvature of PDF around the maximum is related to stability of the system. For highly unstable systems the position PDFs becomes broader as we see from Eq.~\eqref{InvCurvQS}. This equation can be exploited in two ways. Either it can be used to get the local curvature at maximum, $\sigma_{\rm max}^{2}$, from the knowledge of the relaxation rate $\lambda_0$, or, it yields the relaxation rate from measurement of $\sigma_{\rm max}^{2}$ of an experimentally obtained PDF. The local curvature is therefore both measurable and operational characteristic of the system. The recent experiment~\citep{SciRep2017}, already demonstrated Brownian motion in the cubic potential focusing on first-passage properties of the particle~\citep{ARPZRF2016}. Hence the methodology presented here is ready for the experimental test. 

Similar unstable systems should be further analyzed in an underdamped limit, where inertia starts to play an important role leading, e.g.,\ to nonlinear oscillations near the plateau. Such extension is essential since experiments on cooling of nanoparticles in high vacuum has already reached the underdamped regime   \cite{Li2011, Gieseler2013, NovotnyQuidantPRL2016, NovotnyQuidant2017}. Thus, there already exists an experimental platform for probing fundamentals of nonlinear stochastic dynamics in the limit of weak friction. 
 
In recent years, the aforementioned progress in cooling of nanoparticles in optical traps has brought us close to a quantum regime \cite{KieselPNAS2013, NovotnyPRL2016}, where quantum  superposition states can be induced by the cubic nonlinear dynamics \cite{PMarekPRA2011, YukawaPRA2013, MiyataPRA2016}.  Quantum nonlinear effects in the unstable  cubic potential are not only interesting for a fundamental comparison to their stochastic analogs, but they also open doors to quantum simulations and computation with continuous systems \cite{LloydPRL99, GottesmanPRA2001, BartlettPRA2002}.

%%%%%%%%%%%%%%%%%%%%%%%%%%%%%%%%%%%%%%%%%%%%%%%%%%%%%%%%%%%%%%%%%%%%%%%%%%%%%%%%%%%%%%%%%%%%%%%%%%%%%%%%%%%
%%%%%%%%%%%%%%%%%%%%%%%%%%%%%%%%%%%%%%%%%%%%%%%%%%%%%%%%%%%%%%%%%%%%%%%%%%%%%%%%%%%%%%%%%%%%%%%%%%%%%%%%%%%
\section*{Acknowledgments}
%%%%%%%%%%%%%%%%%%%%%%%%%%%%%%%%%%%%%%%%%%%%%%%%%%%%%%%%%%%%%%%%%%%%%%%%%%%%%%%%%%%%%%%%%%%%%%%%%%%%%%%%%%%
%%%%%%%%%%%%%%%%%%%%%%%%%%%%%%%%%%%%%%%%%%%%%%%%%%%%%%%%%%%%%%%%%%%%%%%%%%%%%%%%%%%%%%%%%%%%%%%%%%%%%%%%%%%
R.F.\ and L.O.\ gratefully acknowledge financial support from the Czech Science Foundation (project GB14-36681G). A.R.\ and V.H.\ gratefully acknowledge financial support of the project 17-06716S by the Czech Science Foundation. V.H.\ in addition gratefully acknowledges the support by Alexander von Humboldt foundation. L.O.\ acknowledges the support of the project IGA-PrF-2017-008 by the Palacky University. 

%%%%%%%%%%%%%%%%%%%%%%%%%%%%%%%%%%%%%%%%%%%%%%%%%%%%%%%%%%%%%%%%%%%%%%%%%%%%%%%%%%%%%%%%%%%%%%%%%%%%%%%%%%%
\appendix
\section{Numerical calculation of \texorpdfstring{$Q_{\rm st}(x)$}{Qst}}
\label{sec:app}
%%%%%%%%%%%%%%%%%%%%%%%%%%%%%%%%%%%%%%%%%%%%%%%%%%%%%%%%%%%%%%%%%%%%%%%%%%%%%%%%%%%%%%%%%%%%%%%%%%%%%%%%%%%

The quasi-stationary distribution can be computed as the normalized eigenfunction corresponding to the largest eigenvalue of the Fokker-Planck operator $\mathcal L$, subject to the absorbing boundary condition at $x=a$, cf. Eq.~\eqref{Leqigenvalue}. We have calculated this eigenfunction using the discrete approximation of the generator similar to that used in the recent work \cite{Ratchet2016} for the steady state of a two-dimensional Brownian ratchet.

The main idea is to approximate the exact stochastic process in a semi-infinite continuous state space $(a,\infty)$ by a suitable process on a finite discrete lattice. This is possible because of the strength of the cubic potential for large $|x|$, which allows us to limit the state space to the interval $(a,b)$, with $b \gg (3 k_{\rm B}T/k )^{1/3}$ and $Q_{\rm st}(b) \ll 1$. This is equivalent to keeping the state-space $(a,\infty)$ and redefining the cubic potential $V(x)$ as
\begin{equation}
\tilde{V}(x) = \theta(b - x)V(x) + \theta(x - b) \infty.
\label{eq:VD}
\end{equation}

Let us now discretize the interval $[a,b]$ on $N + 1$ slices of the length $\Delta = (b - a)/N$ and to identify the individual slices with the individual sites of the discrete lattice. We assume that the $i$th site corresponds to the slice next to the point 
\begin{equation}
x(i) = a + \Delta(i - 1),\quad i = 1,\dots,N+1.
\label{eq:x_from_i}
\end{equation}
The vector $\mathbf{p}(t) = (p_1(t), p_2(t), \dots, p_{N+1}(t))$ of probabilities that the discrete system dwells at time $t$ at site $i$ fulfills the master equation 
\begin{equation} 
\frac{d}{dt}\mathbf{p}(t) = L \mathbf{p}(t)
\label{eq:FP_Master}
\end{equation}
with the transition rate matrix $L$, whose the off-diagonal elements are given by
\begin{equation}
l_{ij} = \frac{D}{\Delta^2} \exp\left[-D \frac{\tilde{V}(x(j)) - \tilde{V}(x(i))}{2} \right]
\label{eq:rates}
\end{equation}
and the diagonal elements
\begin{equation}
l_{ii} = -(l_{ii-1} + l_{ii+1}).
\label{eq:RMdiagonal}
\end{equation}
Note that for $i = 1$ we get $r_{11} = -(l_{10} + l_{12})$, but $l_{10}$ is not present elsewhere in the matrix $L$. This is how the absorbing boundary is implemented in the approximate discrete model. Due to this condition the rate matrix no longer fulfills the condition $\sum_{j=1}^{N+1} l_{ij} = 0$ and thus the probability in the Master equation (\ref{eq:FP_Master}) is not conserved, similarly as for the Fokker-Planck equation (\ref{FP}).

 The distribution $P(x,t)$ can be approximately calculated using the formula $P(x(i),t) = p_i(t)/\Delta$ and the approximation becomes exact in the limit $\Delta\to 0$.
Having approximated the generator $\mathcal L$ by the rate matrix $L$, the approximate numerical calculation of the quasi-stationary distribution $Q_{\rm st}(x)$ is a matter of one line of computer code. 

Let us note that the approximation of $Q_{\rm st}(x)$ using this discretization can be very accurate because the matrix $L$ is sparse and thus it is possible to choose very large $N$ (very small $\Delta$). For example, performing the calculation in Matlab on a standard quad-core PC with 16 Gb ram using the command ``sparse'' for constructing the matrix $L$ and the command ``eigs'' for determining the eigenfunction corresponding to the largest eigenvalue of $L$, it is no problem to choose $N$ of the order $10^6$.

%%%%%%%%%%%%%%%%%%%%%%%%%%%%%%%%%%%%%%%%%%%%%%%%%%%%%%%%%%%%%%%%%%%%%%%%%%%%%%%%%%%%%%%%%%%%%%%%%%%%%%%%%%%
\section{Derivation of Eq.~\eqref{FPcond}}
\label{sec:appB}
%%%%%%%%%%%%%%%%%%%%%%%%%%%%%%%%%%%%%%%%%%%%%%%%%%%%%%%%%%%%%%%%%%%%%%%%%%%%%%%%%%%%%%%%%%%%%%%%%%%%%%%%%%%

 To derive dynamical equation~\eqref{FPcond} for the conditioned PDF $Q(x,t)$ we first insert $P(x,t)=Q(x,t)S(t)$ into the Fokker-Planck equation~\eqref{FP} for the unconditioned PDF $P(x,t)$. After dividing the resulting equation by $S(t)$ we obtain
\begin{equation}
\label{B1}
\frac{\partial }{\partial t} Q(x,t) + \frac{Q(x,t)}{S(t)} \frac{d S}{d t}  = \mathcal{L}Q(x,t). 
\end{equation}
Eq.~\eqref{B1} formally differs from  Eq.~\eqref{FP} by the second term on the left-hand side. 
To justify the equation~\eqref{FPcond} we need to identify the conditional probability current~\eqref{CondCurrent} in this second term, i.e., we need to show that 
\begin{equation}
\label{B2}
\frac{1}{S(t)} \frac{d S}{d t} = J_{Q}(a,t). 
\end{equation} 

This is done in two steps. First, we relate the time derivative of the survival probability to the (unconditional) probability current into the absorbing boundary, $d S/ {d t} = J(a,t)$. Here, 
\begin{equation}
J(x,t)= - \left( D \frac{\partial }{\partial x} + \frac{k}{\gamma} x^{2} \right)P(x,t),
\end{equation} 
is the probability current appearing in the generic Fokker-Planck equation~\eqref{FP}, when it is written as the continuity equation~\cite{RiskenBook}, $\partial P/\partial t = - \partial J/\partial x$. Space integration of the continuity equation over the interval $(a,\infty)$ indeed yields $d S/ {d t} = J(a,t)$. 
Second, we divide the relation $d S/ {d t} = J(a,t)$ by the survival probability $S(t)$ and identify $Q(x,t)=P(x,t)/S(t)$ in the expression $J(a,t)/S(t)$, which then is equal to the conditional probability current~\eqref{CondCurrent}, $J(a,t)/S(t) = J_{Q}(a,t)$. This completes the derivation of Eq.~\eqref{B2}and thus of the sought equation~\eqref{FPcond}.

%%%%%%%%%%%%%%%%%%%%%%%%%%%%%%%%%%%%%%%%%%%%%%%%%%%%%%%%%%%%%%%%%%%%%%%%%%%%%%%%%%%%%%%%%%%%%%%%%%%%%%%%%%%
%%%%%%%%%%%%%%%%%%%%%%%%%%%%%%%%%%%%%%%%%%%%%%%%%%%%%%%%%%%%%%%%%%%%%%%%%%%%%%%%%%%%%%%%%%%%%%%%%%%%%%%%%%%
\bibliography{referenzex3}

%%%%%%%%%%%%%%%%%%%%%%%%%%%%%%%%%%%%%%%%%%%%%%%%%%%%%%%%%%%%%%%%%%%%%%%%%%%%%%%%%%%%%%%%%%%%%%%%%%%%%%%%%%%
%%%%%%%%%%%%%%%%%%%%%%%%%%%%%%%%%%%%%%%%%%%%%%%%%%%%%%%%%%%%%%%%%%%%%%%%%%%%%%%%%%%%%%%%%%%%%%%%%%%%%%%%%%%
\end{document}